\documentclass [preprint, prd, aps, nofootinbib]{revtex4}

\begin{document}

\draft
\title{\large \bf The Partition Function and Level Density for
Yang-Mills-Higgs Quantum Mechanics}
\author{\bf S. G. Matinyan$^{(1)}$\footnote{Present address: 3106 Hornbuckle
Place, Durham, NC 27707, USA.  E-mail: sgmatin@aol.com}
and Y. Jack Ng$^{(2)}$\footnote{E-mail: yjng@physics.unc.edu}}
\address{(1) Yerevan Physics Institute, Yerevan, Armenia, 375036\\
(2) Institute of Field Physics, Department of Physics and
Astronomy,
University of North Carolina, Chapel Hill, NC 27599-3255, USA}
%\maketitle

\begin{abstract}

We calculate the partition function $Z(t)$ and the asymptotic integrated
level density $N(E)$ for Yang-Mills-Higgs Quantum Mechanics for two and
three dimensions ($n = 2, 3$).  Due to the infinite volume of the phase
space $\Gamma$ on energy shell for $n= 2$, it is not possible to
disentangle completely the coupled oscillators ($x^2 y^2$-model) from the
Higgs sector.  The situation is different for $n=3$ for which $\Gamma$ is
finite.  The transition from order to chaos in these systems is expressed
by the corresponding transitions in $Z(t)$ and $N(E)$, analogous to the
transitions in adjacent level spacing distribution from Poisson
distribution to Wigner-Dyson distribution.  We also discuss a related 
system with quartic coupled oscillators and two dimensional quartic free
oscillators for which, contrary to YMHQM, both coupling constants are
dimensionless.


\end{abstract}

\maketitle

\bigskip

\section{Introduction}

The discovery of chaoticity in the classical Yang-Mills (YM)
equations\cite{ref1} has attracted attention to the system of coupled
quartic oscillators with the potential $x^2y^2$, where $x$ and $y$ are
functions of time only.  This system is the simplest limiting case for
the homogeneous YM equations (the so-called YM Classical Mechanics) with
$n=2$ degrees of freedom.  Despite its simplicity, the $x^2y^2$ model
exhibits a rich versatile
chaotic behaviour and belongs to the most chaotic
potential systems known.  Not surprisingly, this potential has been used in
many fields, including chemistry, astronomy, astrophysics, and cosmology
(chaotic inflation).

From the quantum mechanical point of view, this system (YM Quantum
Mechanics) has a discrete spectrum\cite{ref2,ref3} in spite of having
an infinite volume of energetically accessible phase space $\Gamma$
\cite{ref4,ref5}
\[
\Gamma = \int dx dy d\dot{x} d\dot{y} \,\,\delta\!\!
\left({1\over2}\dot{x}^2 +
{1\over2}\dot{y}^2
  + {g^2\over2} x^2 y^2 - E\right),
\]
where the dot stands for $d/dt$.  This potential, therefore, violates
the Weil's law\cite{ref6}, a semiclassical relation, which states that the
average number $N(E)$ of quantum energy levels with energy less than
$E$ is
asymptotically proportional to $\Gamma$.

Some time ago, there appeared two important papers\cite{ref7,ref8}
devoted to the calculation of
the partition function $Z(t)$ and the asymptotic integrated level density
$N(E)$ for the $x^2y^2$ model.  Their method was based on an adiabatic
separation in the partition function's dependence on $x$ and $y$ out in
the narrow channels of the equipotential surface $|xy|$ = constant.
Remarkably the dependence on the boundary dividing the two regions
(central ($|x| \leq Q, |y| \leq Q$) and channel
($Q \leq |x|, |y| < \infty$)) with
quite different physics (essentially classical for the first region and
intrinsically quantum for the second region) disappears in the final
result for the partition function.  This insensitivity to the dependence on
the boundary (the value of $Q$) may seem to bode well for this method of
calculation.

However, {\it a priori} we do not expect that the partition function of a
non-integrable system with infinite phase space volume is calculable
without any approximation.  Therefore we seek an alternative approach which
brings to fore the fact that $\Gamma = \infty$ for the $n = 2$ case.
With this in mind, in this paper we calculate the partition function and
the asymptotic integrated level density for the so-called Yang-Mills-Higgs
Quantum Mechanics (YMHQM)\cite{ref9,ref4}.  It is interesting to examine
how $Z(t)$ and $N(E)$ behave in the limit the Higgs coupling to the
$x-$ and $y-$amplitudes vanishes and one recovers the pure
$x^2y^2$-system.
We also consider the YMHQM for the case of $n=3$ for which $\Gamma$ is
finite\cite{ref4,ref5} even for the pure YM system.

\section{Yang-Mills-Higgs Mechanics}

There are several mechanisms that can suppress the classical chaoticity of
chaotic systems with dimension $n = 2$ and, in particular, to the
the YM system (see Ref.\cite{ref5}).  One of them is the well-known Higgs
mechanism.  For spatially homogeneous fields, if only the
interaction of the YM fields with the Higgs vacuum is considered, the
classical Hamiltonian density for $n=2$ is given by
\begin{equation}
H = {1\over2}(\dot{x}^2 + \dot{y}^2) + {g^2\over2} x^2 y^2 +
{v^2\over2}(x^2
+ y^2).
\label{eq1}
\end{equation}

It is known\cite{ref9} that there is a classical ``phase transition'' from
chaos to regular motions as the vacuum expectation value of the Higgs
field $\langle \phi \rangle = v$ gets large enough.  Here there is
one control
parameter $\kappa = {g^2v^4\over H}$, where $H$ is the conserved
energy density.  At large $\kappa$ the motion is regular, whereas at
$\kappa = 0$ one deals with the developed chaos of YM Classical Mechanics.
In fact, chaos appears already at $\kappa \approx 0.60$.\cite{ref9}
Therefore, in the study of the quantum counterpart of Eq. (\ref{eq1}), we
expect that there is a transition from one type of $N(E)$ to another,
depending on the parameter $v$.  The analogous transition in the
adjacent energy level spacing distribution, as a function of $v$, was
predicted\cite{ref10} and established in several papers\cite{ref11}.

In the next section we calculate the partition function for the
quantum-mechanical counterpart of the classical Hamiltonian density
Eq. (\ref{eq1}).  In section IV, we calculate the integrated level density.
Section V is devoted to the $n=3$ case.  We conclude in the last section with
some discussions involving a related system with potential of
the form ${g^2 \over 2}x^2y^2 + {b^2\over4}(x^4 + y^4)$.

\section{Partition function for YMH quantum mechanics}

The quantum Hamiltonian corresponding to Eq. (\ref{eq1}) is given by (mass
$m=1$ in the following)
\begin{equation}
\mathcal{H} = {\overrightarrow{{\bf p}}^2 \over2} + {g^2 \over 2} x^2y^2 +
   {v^2 \over 2} (x^2 + y^2),
\label{eq2}
\end{equation}
where $\overrightarrow{{\bf p}} = -i \hbar \overrightarrow{{\bf \nabla}}$
is
the momentum operator.  First a word on units.  All quantities below are
given in units of energy $E$: $[\mathcal{H}] = E$, $[t] = E^{-1}$, $[x] = [y] =
E^{1/4}$, $[g] = E^0$,
$[v] = E^{1/4}$, $[\hbar] = E^{3/4}$.  It is obvious that the
operator Eq. (\ref{eq2}) has a discrete spectrum.

Here we calculate the trace of the heat kernel $\exp (- t \mathcal{H})$, the
partition function for the Hamiltonian operator.  The trace is defined for
any quantum operator $\mathcal{A}_W$ in the Wigner representation (i.e.,
in the classical 2n-dimensional phase space)
\begin{equation}
Tr( \mathcal{A}_W) = {1\over (2 \pi \hbar)^n} \int d
\overrightarrow{{\bf p}}
  d \overrightarrow{{\bf q}} A_W (\overrightarrow{{\bf p}},
\overrightarrow{{\bf q}}).
\label{eq3}
\end{equation}
Thus, we have
\begin{equation}
Z(t) = Tr (e^{-t \mathcal{H}}) = \int_0^\infty dE e^{-tE} \rho(E).
\label{eq4}
\end{equation}
The second equation expresses the fact that the partition function $Z(t)$
and the density of eigenstates $\rho(E)$ form a Laplace transform pair.
The asymptotic integrated density of states $N(E)$ is given
by the inverse Laplace transform of $Z(t)/t$
\begin{equation}
N(E) = \int_0^E dE' \rho(E') = L^{-1}\! \left({Z(t)\over t}\right).
\label{eq5}
\end{equation}

Integrating Eq. (\ref{eq4}) over $p_x$ and $p_y$ for the Hamiltonian
given by Eq. (\ref{eq2}), we get
\begin{equation}
Z(t) = {1\over (2 \pi \hbar)^2} {2 \pi \over t}
  \int dx dy e^{- t [v^2 (x^2 + y^2) + g^2 x^2y^2] / 2}.
\label{eq6}
\end{equation}
After the integration over $x$, we can make a change of variable from $y$
to $s = {1\over2}t v^2 y^2 $ to obtain
\begin{equation}
Z(t) = {1\over (2 \pi \hbar)^2} \left( {2 \pi \over t}\right)^{3/2}
  \sqrt{{2 \over t v^4}}\int_0^\infty ds e^{-s} s^{-1/2}
  \left( 1 + {s\over 2z} \right)^{- 1/2},
\label{eq7}
\end{equation}
with $z \equiv {t v^4\over 4 g^2}$.  But aside from a factor of $\sqrt{2z}
e^z$, the integral is
just a representation of $K_0(z)$, the modified Bessel
function of the third kind or the MacDonald function of order 
$0$.
(see, e.g., p.140 of Ref.\cite{ref12}).  Thus we arrive at the precise
expression for the partition function corresponding to the Hamiltonian
Eq. (\ref{eq2})
\begin{equation}
Z(t) = {1\over \sqrt{2 \pi}} {1\over g \hbar^2 t^{3/2}}
  \,\exp\!\left({tv^4\over
  4 g^2}\right) K_0\!\left({tv^4 \over 4 g^2}\right).
\label{eq8}
\end{equation}
So far we have not made any calculational approximations.
For two uncoupled oscillators (corresponding to $g=0$) Eq. (\ref{eq8})
gives (with the aid of $K_0 (z) \simeq \sqrt{{\pi \over 2z}} e^{-z}$
as $z \rightarrow \infty$)
\begin{equation}
Z(t) = {1 \over (\hbar v t )^2},
\label{eq9}
\end{equation}
as it should.  For the sake of completeness, let us also give the
partition function for one oscillator ($n=1$)
\addtocounter{equation}{-1}
\renewcommand{\theequation}{\arabic{equation}$'$}
\begin{equation}
Z(t) = {1\over \hbar v t}.
\label{XX}
\end{equation}
\renewcommand{\theequation}{\arabic{equation}}

From Eq. (\ref{eq8}), it is obvious that $Z(t)$ diverges logarithmically in
the limit $v \rightarrow 0$ (the $x^2y^2$ model).  Indeed, using
$K_0(z) \simeq -\log(z/2) - C$ (with $C$ being the Euler constant) as $z
\rightarrow 0$, we get, for $v \simeq 0$,
\begin{equation}
Z(t) \simeq {1\over \sqrt{2 \pi}} {1\over g \hbar^2 t^{3/2}}
  \left( \log{8 g^2 \over t v^4} - C \right).
\label{eq10}
\end{equation}
The impossibility to disentangle the coupled oscillators from the uncoupled
ones is a reflection of the logarithmic divergence of the phase space
volume $\Gamma$ on the $E$-shell at $v=0$ for $n=2$, as
we have mentioned above.
For $n=3$ (see section V below), $\Gamma$ is finite and the precise $v=0$
limit exists.

\section{Integrated level density $N(E)$ ($n=2$)}

It is well-known that the asymptotic integrated density of states $N(E)$ is
related to the small
$t$ divergence of the partition function $Z(t)$, according
to the Karamata-Tauberian theorem (see, e.g., Ref. \cite{ref3}).  Using Eq.
(\ref{eq8}) and Eq. (\ref{eq5}), we obtain the precise expression for
$N(E)$ (with the aid of formula 3.16.3.3 of Ref.\cite{ref13})
\begin{eqnarray}
N(E)&=& {1\over \sqrt{2 \pi}} {1\over g \hbar^2}\, L^{-1}\! \left( {1\over
t^{5/2}}
\,\exp\!\left({t v^4\over 4 g^2}\right)
K_0\! \left({t v^4 \over 4 g^2}\right)\right)\nonumber\\
&=& {E^2 \over 2 \hbar^2 v^2}\, F\!\left({1\over2},{1\over 2};3;-{2g^2 E \over
v^4}\right),
\label{eq11}
\end{eqnarray}
where F is the Gauss hypergeometric function.  For $g=0$ (the case of two
uncoupled oscillators) we have, from Eq. (\ref{eq11}),
\begin{equation}
N(E) = {1\over 2 \hbar^2 v^2} E^2,
\label{eq12}
\end{equation}
in agreement with Eq. (\ref{eq9}).  For completeness, for one oscillator,
we have, from Eq. (\ref{XX}),
\addtocounter{equation}{-1}
\renewcommand{\theequation}{\arabic{equation}$'$}
\begin{equation}
N(E) = {1\over \hbar v} E.
\label{XXX}
\end{equation}
\renewcommand{\theequation}{\arabic{equation}}

For the $v \rightarrow 0$ limit, one needs the asymptotic expression of
$F(a,b;c;z)$. Using formula 9.7.7 of Ref.\cite{ref12}, we get
\begin{equation}
N(E) \simeq {2 \sqrt{2} \over 3 \pi} {E^{3/2}\over g \hbar^2}
\left( \log{g^2 E \over v^4} + 5\log2 - {8\over3} \right).
\label{eq13}
\end{equation}
One can verify that the $Z(t)$ and $N(E)$ for
Eq. (\ref{eq9}), Eq. (\ref{XX}), and the corresponding Eq. (\ref{eq12}),
Eq. (\ref{XXX}), and the {\it ratio} of
the logarithmic parts of Eq. (\ref{eq10}) and
Eq. (\ref{eq13}), are in accordance with Theorems
1.1 and 1.4 of Ref. \cite{ref3}.  Also
from Eq. (\ref{eq12}) and Eq. (\ref{eq13}),
we see that the transition from
order ($g=0$) to chaos ($g\neq 0$ and large, $v$ small, $\kappa
\neq 0$)\footnote{We need to emphasize that, from the results of
Ref.\cite{ref9}, the chaotic regime for $N(E)$ actually begins at
finite $v = (\kappa E / g^2)^{1/4}$.}
corresponds to the change in the $E$-dependence of
$N(E)$ from $N(E) \sim E^2$ to $N(E) \sim E^{3/2} \log E$, analogous to
the change in the neighbour level spacing distribution from the Poisson
distribution to the Wigner-Dyson distribution.  It can be seen that
as the power $\alpha$ of the homogeneous potential $|xy|^\alpha$
increases from 1 to $\infty$,
there is \cite{ref3} a systematic decrease in the power of
$E$ in $N(E)$ from 2 to 1, the latter corresponding to the case of the
hyperbola billiard with $N(E) \sim E \log E$ (see Ref.\cite{ref14}).
This observation may be relevant to the Hilbert-Polya-Berry program 
in identifying a quantum (chaotic) Hamiltonian whose eigenvalues 
reproduce the Riemann zeta-function zeros.

\section{Some remarks on the three dimensional YM and YMH quantum
mechanics}

Let us now consider the YMH Quantum Mechanics for $n=3$ with the
Hamiltonian (with mass $m = 1$ again)
\begin{equation}
\mathcal{H} = {1\over2} \overrightarrow{{\bf p}}^2 +
 {g^2 \over 2}(x^2y^2 + y^2z^2 +z^2x^2) + {v^2\over2} (x^2 + y^2 + z^2).
\label{eq14}
\end{equation}
For the partition function, the integrations over $\overrightarrow
{{\bf p}}$ can be easily carried out.  Introducing the cylindrical
coordinates ($x = r\cos \phi, y = r\sin \phi, z = z$) and performing the
integrations over $z$ and $\phi$, we obtain
\begin{equation}
Z(t) = {1\over (2 \pi \hbar)^3} \left({2 \pi \over t} \right)^2 2 \pi
  \int_0^\infty {r dr \over \sqrt{v^2 + g^2r^2}}
  \,\exp\!\left(-{t v^2r^2 \over 2} - {tg^2r^4 \over 16}\right)
  I_0\!\left({tg^2r^4 \over 16}\right),
\label{eq15}
\end{equation}
where $I_0(z)$ is the modified Bessel function of the first kind of order
$0$.

We see that
the integral in Eq. (\ref{eq15}) is well behaved at $v = 0$; it is equal to
\addtocounter{equation}{-1}
\renewcommand{\theequation}{\arabic{equation}$'$}
\begin{equation}
{1\over 4g} \int_0^\infty du {1\over u^{3/4}} e^{-g^2 t u / 16}
  I_0\!\left({g^2 t u \over 16}\right).
\label{XXXX}
\end{equation}
\renewcommand{\theequation}{\arabic{equation}}
The integration over $u$ can be done by using formula 2.15.3.3 of
Ref.\cite{ref15} to yield
\begin{equation}
Z(t) = {\Gamma^3({1\over4}) \over 2^{7/4} \pi^{3/2}}{1\over g^{3/2} 
\hbar^3 t^{9/4}},
\label{eq16}
\end{equation}
which agrees with Eq. (4.5) of Ref.\cite{ref8} (after taking into account
the extra factor of ${1\over2}$ in the definition of the potential in Eq.
(\ref{eq14})).  The difference between the $n = 2$ and $n = 3$ cases is due
to the fact that for $n = 3$ the energetically accessible region pinches as
$1/x^2$ at large $x$, in contrast to the $n = 2$ case for which it pinches
as $1/x$,
leading to the logarithmic divergent result (Eq. (\ref{eq10})) at $v = 0$.
Note that the contibution from the channels ($t^{1/4}x >> 1$) for $n = 3$
is negligible in the method of separating the central part from the
channels used in Ref.\cite{ref7,ref8}.  Thus for $n = 3$ the coupled
oscillators are disentangled from the free oscillators.  From
Eq. (\ref{eq16}), it follows that
\begin{equation}
N(E) = {16\over45}{2^{1/4} \Gamma^2({1\over4}) \over \pi^{3/2}}
\left({E^{3/4} \over \sqrt{g}
\hbar}\right)^3,
\label{eq17}
\end{equation}
which agrees with the result given in Ref.\cite{ref8} (the first of Eq.
(5.19)).

Finally, for the sake of completeness, here are the expressions for
$Z(t)$ and $N(E)$ at $g = 0$ for the case of $n = 3$.  Eq. (\ref{eq15})
yields
\begin{equation}
Z(t) = {1\over (\hbar v t)^3}.
\label{eq18}
\end{equation}
Of course, Eq. (\ref{eq9}), Eq. (\ref{XX}), and Eq. (\ref{eq18}) are
in accordance
with the well-known relation of the trace of the n-harmonic-oscillator
heat kernel, viz., $Z(t) = \left( 2\sinh{\hbar v t\over
2}\right)^{-n}$ in the quasiclassical approximation ($t \rightarrow
0$)\cite{ref7,ref8}.  From Eq. (\ref{eq18}) we have, for $N(E)$,
\begin{equation}
N(E) = {1\over 6 \hbar^3 v^3} E^3.
\label{eq19}
\end{equation}
Eq. (\ref{eq16}) - Eq. (\ref{eq19}) show that the $Z(t)$ and $N(E)$
for the $n=3$ case are in full agreement with Theorem 1.1 of
Ref.\cite{ref3}.

\section{Discussions}

From the previous two sections, we see that the $\log E$ 
factor in $N(E)$ is specific to the
$x^2y^2$-model (with or without the Higgs term), and is not generic 
to chaotic systems with potentials
$|xy|^\alpha$ ($\alpha = 2,3,...$)
for $n > 2$.  But one may wonder whether it is generic to other 
homogeneous potentials for $n = 2$.  The answer depends on the type 
of homogeneous potentials added to the $x^2y^2$ term.  Consider the
following example of 
$n=2$ system (with finite phase space), described by
the following quantum Hamiltonian
\begin{equation}
\mathcal{H} = {\overrightarrow{{\bf p}}^2 \over2} + {g^2 \over 2} x^2y^2 +
   {b^2 \over 4} (x^4 + y^4).
\label{eq20}
\end{equation}
Like $g$, $b$ is dimensionless.  It is known that for $b^2 = g^2/3$ the 
system Eq. (\ref{eq20}) is integrable.\cite{ref16}
In the calculation of $Z(t)$, the 
integrations over $x$ and $y$ can best be done by using the 
polar coordinates.
After the gaussian integration over 
the radial coordinate, 
the integration over the angle yields the complete elliptic
integral of the first kind\cite{ref12} which is related to the Gauss
hypergeometric function.  The result is given by
\begin{equation}
Z(t) = {\sqrt{\pi}\over2} {1\over b \hbar^2 t^{3/2}} F\!\left(
  {1\over2},{1\over2};1;{b^2 - g^2 \over 2 b^2}\right).
\label{eq21}
\end{equation}
The corresponding asymptotic integrated density 
of states $N(E)$ is given by
\begin{equation}
N(E) ={2 E^{3/2} \over 3 b \hbar^2} F\!\left(
  {1\over2},{1\over2};1;{b^2 - g^2 \over 2 b^2}\right).
\label{eq22}
\end{equation}

For the $b \rightarrow 0$ limit, we use formula 9.7.7 of Ref.\cite{ref12} 
again to get
\begin{equation}
Z(t) \simeq {1\over \sqrt{2 \pi}}{1\over g \hbar^2 t^{3/2}}\,
  \log{8 g^2 \over b^2}.
\label{eq23}
\end{equation}
We note the logarithmic divergence as $b \rightarrow 0$, i.e., when
one tries to disentangle the coupled quartic oscillators from the free
quartic oscillators.  A simple inverse Laplace transform yields the level
density
\begin{equation}
N(E) \simeq {2 \sqrt{2}\over 3 \pi}{E^{3/2}\over g \hbar^2}\,
  \log{8 g^2 \over b^2}.
\label{eq24}
\end{equation}
Just as there is no $\log t$ dependence in $Z(t)$, we 
find no $\log E$ dependence in $N(E)$ for this system.  This 
result is not surprising from the viewpoint 
of dimensional considerations since both couplings in this example are 
dimensionless.  In the case of YMHQM, the coupling $v$ is dimensionful, 
and due to this, $\log t$ and $\log E$ appear in Eq. (\ref{eq10}) and 
Eq. (\ref{eq13}) respectively in the $v \rightarrow 0$ limit. 

For completeness we also give the 
partition function for the $g = 0$ case which corresponds to an
integrable system. 
Using 
$F({1\over2},{1\over2};1;{1\over2}) = \Gamma^2({1\over4})/ (2 \pi^{3/2})$,
we have
\begin{equation}
Z(t) = {\Gamma^2({1\over4}) \over 4 \pi}{1\over b \hbar^2 t^{3/2}}.
\label{eq25}
\end{equation}
The corresponding asymptotic integrated density of states is given by
\begin{equation}
N(E) = {\Gamma^2({1\over4}) \over 3 \pi^{3/2}}
  {E^{3/2}\over b \hbar^2}.
\label{eq26}
\end{equation}
For another integrable case 
with $b^2 = g^2 /3$ in this example\cite
{ref16}, using $F({1\over2},{1\over2};1;-1) = 
\Gamma^2({1\over4})/ (2 \pi)^{3/2}$, we get
\begin{equation}
Z(t) = {\Gamma^2({1\over4}) \over 2^{5/2} \pi}
{1\over b \hbar^2 t^{3/2}},
\label{eq27}
\end{equation}
and
\begin{equation}
N(E) = {\Gamma^2({1\over4})\over 3 (2 \pi^3)^{1/2}}
 {E^{3/2}\over b \hbar^2}.
\label{eq28}
\end{equation}
Finally, for the third integrable case with $b = g$ we immediately obtain
the $Z(t)$ and $N(E)$ from Eq. (\ref{eq21}) and Eq. (\ref{eq22}) using
$F({1\over2},{1\over2};1,0) = 1$.

\bigskip

\section*{Acknowledgements}

This work was supported in part by the US Department of Energy
and by the Bahnson Fund of University of North
Carolina at Chapel Hill.  We thank Berndt Muller for his interest in this 
work, and 
Matthias Brack for helpful correspondence and for calling our attention to
related work given in Ref. 17.

\end{document}